\documentclass{article}
\usepackage[utf8]{inputenc}
\usepackage{mathtools}
\usepackage[table]{xcolor}
\usepackage{amssymb,graphicx}
\usepackage{epstopdf}
\usepackage{amsmath,amsfonts}
\usepackage{epsfig}
\usepackage{xcolor}
\usepackage[T1, T2A]{fontenc}
\usepackage[english]{babel}
\usepackage{float}
\usepackage{soul}
\usepackage{graphicx}
\usepackage{subcaption}
\usepackage{geometry}
\usepackage{hyperref}
\usepackage{psfrag}
\usepackage{slashed}
\usepackage{braket}
\usepackage{bm}
\usepackage{siunitx}
\usepackage{soul}
\begin{document}
\begin{flushright}
INR-TH-2025-015
\end{flushright}
\vspace{10pt}
\begin{center}
  {\LARGE \bf On the link between finite QFT and standard RG approaches} \\
\vspace{20pt}
Y. A. Ageeva$^{a,b,c}$\footnote[1]{{\bf email:}
    ageeva@inr.ac.ru} and A. L. Kataev$^{a,d}$\footnote[2]{{\bf email:}
    kataev@ms2.inr.ac.ru}\\
\vspace{15pt}
  $^a$\textit{
Institute for Nuclear Research of
         the Russian Academy of Sciences,  60th October Anniversary
  Prospect, 7a, 117312 Moscow, Russia }\\
  $^b$\textit{
Institute for Theoretical and Mathematical Physics,
  M.V.~Lomonosov Moscow State University, Leninskie Gory 1,
119991 Moscow,
Russia }\\
  $^c$\textit{
Department of Particle Physics and Cosmology, Physics Faculty, M.V. Lomonosov Moscow State University, Leninskie Gory 1-2, 119991 Moscow, Russia }\\
  $^d$\textit{
Bogoliubov Laboratory of Theoretical Physics,
Joint Institute for Nuclear Research,
141980, Dubna, Russia }\\

    \end{center}
    \vspace{5pt}
\begin{abstract}
A finite formulation of quantum field theory based on a system of differential equations reminiscent of the Callan–Symanzik equations is discussed. This system of equations was previously formulated in the bare language. We rederive it in a fully renormalized language. For the latter, within a simple 
$\phi^4$ toy model, it is shown that with a specific choice of renormalization conditions -- namely, the on‑shell scheme for the renormalized mass -- this class of finite renormalization prescriptions is equivalent to the standard renormalization‑group equation written in the Callan--Symanzik--Ovsyannikov form.
\end{abstract}
\newpage
\section{Introduction}
\label{sec:intro}

In the standard approach to quantum field theory (QFT), one encounters divergent multi‑loop Feynman diagrams. However, after the procedure of regularization to extract UV divergences and subsequent renormalization the final answer for the physical observables becomes finite~\cite{Bogolyubov:1959bfo,Itzykson:1980rh,Collins:1984xc}. 

Nevertheless, it is also possible to think about the whole way of computations in QFT as follows: whatever calculation procedure we use, ultimately it is a \textit{map} between finite parameters characterizing the theory and finite Green’s functions. From this point of view, it looks pretty natural to consider such an approach, where one does not meet any divergent expressions at any stage of the computation.

There are several approaches in QFT, in which intermediate divergences are handled carefully. One such approach is the Bogolubov--Parasuk--Hepp--Zimmermann (BPHZ) renormalization procedure \cite{Bogoliubov:1957gp,Hepp:1966eg,Zimmermann:1969jj}  with the application of the R-operation. It is known that this method is perfectly  applicable for renormalizable theories. Moreover, recently there have been works \cite{Kazakov:2020kbj,Kazakov:2020mfp, Kazakov:2023ugc} on how this method can be used for non-renormalizable theories as well.  

We focus our attention on another formulation, which is based on \textit{the system of differential equations, which are reminiscent of Callan--Symanzik equations} \cite{Callan:1970yg,Symanzik:1970rt}. 
This approach was first formulated in Refs.~\cite{Blaer:1974foy,Callan:1975} as a proof of validity of the multiplicative renormalization procedure. The idea of this procedure corresponds \textit{to the differentiation} of bare field propagator
with respect \textit{to mass}. Application of this operation reduces the degree of divergence of the particular graph.
It has been shown that within this program one can compute quantum corrections to 
$n$-point Green’s functions and to the effective potential in a finite manner, i.e. no intermediate divergences arise \cite{Mooij:2021ojy,Mooij:2021lbc,Mooij:2024rys,Ageeva:2024qie}.
Below, we refer to this method and to the associated system of differential equations as the ``CS method'' and the ``CS equations'', respectively.

This work is devoted to the study of a subtlety related to this finite CS method. By construction, the CS equations were formulated in a ``bare'' language. The latter particularly may resemble a specific choice or redefinition of renormalization factors at each order of perturbation theory as it was considered, for example in Ref.~\cite{Kataev:2013eta}, for the charge renormalization factor, $Z_3 = 1$, in supersymmetric QED, or the specific finite renormalization procedure in $d=4$ spacetime \cite{Lenshina:2020edt}.  
An interesting question arises: whether it is possible to rederive the CS equations in a fully renormalized language, explicitly introducing the renormalization of the field, the mass, the coupling constant, and all correlation functions. The answer is ``yes''. In this paper, such a derivation is done for a simple and clear $\phi^4$ toy model. Moreover, this rederivation shows that \textit{under certain conditions which are related to the choice of  concrete renormalization scheme -- on‑shell scheme (OS) for the renormalized mass -- the system of the CS equations is equivalent to the known equation of the renormalization group (RG)} which is written in the Callan--Symanzik--Ovsyannikov form \cite{Callan:1970yg,Symanzik:1970rt,Ovsyannikov:1956fa}. 

As it was mentioned above, CS equations was first formulated as a proof of validity of the multiplicative renormalization procedure \cite{Blaer:1974foy,Callan:1975}. Indeed, the proof constructs the renormalization‑group functions and renormalized correlation functions order by order without encountering divergences. It is important to note, that considerations of Refs.~\cite{Blaer:1974foy,Callan:1975} explicitly make use of the normalization conditions. However, the authors of Refs.~\cite{Naud:1998yg,Carvalho:2013jai} have shown, that it is possible to prove the renormalizability within the dimensional regularization \cite{tHooft:1972tcz} minimal
subtraction (MS) \cite{tHooft:1973mfk} scheme. In this paper we do not consider this way. Our work does not attempt to prove renormalizability; instead, it shows that the CS equations and the RG equation are equivalent, but only in the on‑shell mass‑renormalization scheme.

This paper is organized as follows. In section \ref{sec:general} we specify the model to work with and introduce all necessary notations. Section \ref{sec:compare} rederives the CS equations in terms of renormalized quantities within the chosen $\phi^4$ toy model; in section \ref{sec:on-shell} we discuss the equivalence of these equations to the standard RG equation in the Callan--Symanzik--Ovsyannikov form, clarifying the renormalization‑scheme conditions under which this equivalence holds. Section \ref{sec:concl} concludes.

\section{Generalities}
\label{sec:general}
In this short section, we discuss the chosen setup and notation. To begin with, let us specify the model we are going to consider. For simplicity, we choose standard $\phi^4$ toy model:
\begin{align}
\label{L}
    &\mathcal{L} = \frac{1}{2}\partial_{\mu}\phi_0\partial^{\mu}\phi_0 - \frac{m_0^2}{2}\phi_0^2 + \frac{\lambda_0}{4!}\phi_0^4 ,
\end{align}
where the subscript $0$ denotes bare quantities. The signature of the metric is $(+,-,-,-)$\footnote{The results of Refs.\cite{Mooij:2021ojy,Mooij:2021lbc,Mooij:2024rys,Ageeva:2024qie} to which we refer were obtained in mostly positive signature $(-,+,+,+)$ as well as with the minus sign in the interaction term, i.e. $-\frac{\lambda_0}{4!}\phi_0^4$. That is why, the propagator in Refs.\cite{Mooij:2021ojy,Mooij:2021lbc,Mooij:2024rys,Ageeva:2024qie} is $\frac{-i}{p^2+m_0^2}$ and the sign in the Feynman rule for corresponding vertex is $-i\lambda_0$. }.

Later on, we are going to consider equations for  \textit{n-point one-particle irreducible (1PI) correlation functions}. To introduce the latter, we first recall the definition of Green's function $G_{n}$:
\begin{align}
\label{green}
    &(2\pi)^4\delta^{(4)}(p_1+\ldots+p_n)G_{n} (p_1,\ldots,p_{n}) = \int \prod^n_{i=1}
    d^4x_i e^{-i p_ix_i}\braket{\phi_0(x_1)\ldots\phi_0(x_n)},
\end{align}
for $n$ external momenta. The notation $\braket{\ldots}$ denotes the time‑ordered product of fields, and only connected graphs are included. Next, to obtain 1PI Green's function $G^{\text{1PI}}_{n}(p_1,\ldots,p_n)$, one should take in eq.~\eqref{green} only diagrams that cannot be divided into two disconnected parts by cutting any internal line. Removing the external propagators (i.e., amputating the external legs) yields $\Gamma^{(n)}$:
\begin{align}
\label{Gamma_n_def}
    \Gamma^{(n)}(p_1,\ldots,p_n) \equiv  \prod^n_{i=1}\Big(\frac{p^2-m_0^2}{i}\Big) G_n^{\text{1PI}}(p_1,\ldots,p_n).
\end{align}
For example, the two-point Green's function at tree level reads
\begin{align}
    G^{\text{tree}}_2(p) = \frac{i}{p^2-m_0^2}.
\end{align}
Hence $\Gamma_{\text{tree}}^{(2)}(p)$ is just the inverse propagator
\begin{align}
    \Gamma_{\text{tree}}^{(2)}(p) \equiv G_2^{\text{1PI}}(p)\times \Big(\frac{p^2-m_0^2}{i}\Big)^2 = -i(p^2-m_0^2).
\end{align}
One can also fix four-point 1PI correlation function in the theory \eqref{L} at tree level:
\begin{align}
    \Gamma_{\text{tree}}^{(4)} = i\lambda_0.
\end{align}

Above we have considered the expressions for bare objects only. It is known, that one loop and higher corrections to these bare correlation functions, etc include divergent Feynman integrals. Thus, for our further purposes, we now introduce the following relations between renormalized and bare correlation functions:
\begin{align}
    &\Gamma_0^{(n)} (\lambda_0,x_{m_0},\{p_i/\mu\}) = Z_n(\lambda,x_m) \Gamma^{(n)} (\lambda,x_m,\{p_i/\mu\})\label{G2 bold renor},
\end{align} 
where $x_{m_0}\equiv m_0/\mu$, $x_{m}\equiv m/\mu$ and where $\lambda_0$, $m_0$ are bare coupling and mass; $\mu$ is an arbitrary parameter, related to the renormalization procedure, and which can be also used to fix the procedure of subtracting UV (sub)divergencies. We also have the notation for $n$ external momenta $\{p_i/\mu\} \equiv (p_1/\mu, p_2/\mu,\ldots,p_n/\mu)$.

For now we turn to the full (not only tree) correlation function. The renormalization constant $Z_n$ is constructed so that it includes all divergencies in the regularization parameter.  Thus $\Gamma^{(n)} (\lambda,x_m)$ is finite renormalized correlation function. The relations between bare and renormalized field, coupling and mass are
\begin{subequations}
\label{eq:all renorm}
\begin{align}
    &\phi_0 = Z_{\phi}^{1/2} \phi,\\
    &\lambda_0 = Z_{\lambda} \lambda,\\
    &m_0 = Z_{m} m,
\end{align}
\end{subequations}
where $Z_{\phi,\lambda,m} = Z_{\phi,\lambda,m}(\lambda,x_m)$.
Defining the renormalization conditions \eqref{G2 bold renor} and \eqref{eq:all renorm}, one has
\begin{align}
\label{eq:Z2 Z4}
    Z = Z_2, \quad Z_{\lambda} = \frac{Z_4}{Z_2^2} ,
\end{align}
where $Z_2$ and $Z_4$ are defined by \eqref{G2 bold renor} with $n=2,4$, respectively. 

The calculation of the renormalization factors $Z_2$, $Z_4$ is performed in a specific renormalization scheme. Physical observables ultimately cannot depend on the choice of the particular scheme. This requirement imposes certain constraints on the parameters defined in different renormalization programs and finally leads to the RG approach, see Refs.~\cite{Bogolyubov:1959bfo,Itzykson:1980rh,Collins:1984xc} for details. 
The related RG equation for the n-point correlation function has the following form \cite{Itzykson:1980rh}
\begin{align}
\label{RG}
    \Big[\mu \frac{\partial}{\partial \mu} + \beta(\lambda,x_m) \frac{\partial}{\partial \lambda} + \gamma_m(\lambda,x_m) m^2 \frac{\partial}{\partial m^2} + n \gamma_n(\lambda,x_m) \Big] \Gamma^{(n)}(\lambda,x_m,\{p_i/\mu\}) = 0,
\end{align}
also known as Callan–Symanzik–Ovsyannikov differential RG equation, where within the perturbation theory with the fixed bare parameters $\lambda_0$ and $x_{m_0}$
\begin{subequations}
\label{eq:RG betas}
\begin{align}
    &\beta(\lambda,x_m) \equiv \mu \frac{\partial\lambda}{\partial\mu}\Big|_{x_{m_0},\lambda_0=\text{fixed}} = \sum_{k\geq 0}\beta_k (x_m)\lambda^{k+2} ,\\
    \label{gamma m RG}
    &\gamma_m(\lambda,x_m) \equiv \mu\frac{\partial \text{ln}\;m^2}{\partial \mu}\Big|_{x_{m_0},\lambda_0=\text{fixed}} = \sum_{k\geq 0}\gamma_{m,k} (x_m)\lambda^{k+1},\\
    &\gamma_n(\lambda,x_m) \equiv \mu \frac{\partial  \text{ln}Z_n}{\partial \mu}\Big|_{x_{m_0},\lambda_0=\text{fixed}} = \sum_{k\geq 0}\gamma_{n,k} (x_m)\lambda^{k+1},
\end{align}
\end{subequations}
where  the dependence of $\beta_k(x_m)$, $\gamma_{m,k} (x_m)$, and $\gamma_{n,k} (x_m)$ on $x_m$ arises in general in the arbitrary (not MS-like) renormalization schemes; also, $\lambda = \lambda(\mu)$. Let us also note, that $\beta_k(x_m)$, $\gamma_{m,k} (x_m)$, and $\gamma_{n,k} (x_m)$ are directly evaluated through the set of renormalized Feynman diagrams.  As usual, $\beta$ is $\beta$-function, $\gamma_m$ is anomalous dimension of mass and $\gamma_n$ anomalous dimension of the field, respectively. We emphasize once again that these notations \eqref{eq:RG betas} are introduced with the fixed bare $m_0,\lambda_0$. 

The particular form of equation \eqref{RG} and all expressions  \eqref{eq:RG betas} depends on how we define the renormalization prescriptions, which encoded in the following general expression
\begin{align}
\label{R-op}
    & \Gamma^{(n)} (\lambda,x_m,\{p_i/\mu\}) = (1 - \mathcal{K}\mathcal{R}')\Gamma_0^{(n)} (\lambda_0,x_{m_0},\{p_i/\mu\}) \equiv  Z_n^{-1}(\lambda,x_m)\Gamma_0^{(n)} (\lambda_0,x_{m_0},\{p_i/\mu\}). 
\end{align}
Here $\mathcal{R}'$-subtracts the divergent subgraphs and the $\mathcal{K}$-operation reflects the overall remaining divergencies. For instance, in the MS-scheme the divergences are isolated as poles in the dimensional regulator $\epsilon$, thus $\mathcal{K}$ will return only these poles; see Refs.~\cite{Bogolyubov:1959bfo,Kazakov:2008tr} for more details about $\mathcal{R}$-operation itself and its application in the MS-scheme in Refs.~\cite{Caswell:1981ek,Chetyrkin:2017ppe,Herzog:2017bjx}. 

Let us now consider the RG-like equations, related to the case when renormalized parameters fixed. In this case, instead of the prescription \eqref{R-op}, one should use the following expression:
\begin{align}
\label{R-op-new}
    & \Gamma^{(n)} (\lambda,x_m,\{p_i/\mu\}) = (1 - \mathcal{K})\Gamma_0^{(n)} (\lambda_0,x_{m_0},\{p_i/\mu\}) \equiv  \tilde{Z}_n^{-1}(\lambda,x_m)\Gamma_0^{(n)} (\lambda_0,x_{m_0},\{p_i/\mu\}).
\end{align} 
Note that here the action of $\mathcal{R}' = 1$ operation presumes that the subdivergencies of $\Gamma_0^{(n)} (\lambda_0,x_{m_0},\{p_i/\mu\})$ are not subtracted. 
In this case, the corresponding equation has the form:
\begin{align}
\label{RG zero}
    \Big[\mu \frac{\partial}{\partial \mu} + \tilde{\beta}(\lambda_0,x_{m_0}) \frac{\partial}{\partial \lambda_0} + \tilde{\gamma}_m(\lambda_0,x_{m_0}) m_0^2 \frac{\partial}{\partial m_0^2} + n \tilde{\gamma}_n(\lambda_0,x_{m_0}) \Big] \Gamma^{(n)}(\lambda_0,x_{m_0},\{p_i/\mu\}) = 0.
\end{align}
Here the bare RG-functions are defined with fixed renormalized parameters $\lambda$, $x_m$ in the following form
 \begin{subequations}
\label{eq:RG betas zero}
\begin{align}
    &\tilde{\beta}(\lambda_0,x_{m_0}) \equiv \mu \frac{\partial\lambda_0}{\partial\mu}\Big|_{x_{m},\lambda=\text{fixed}} = \sum_{k\geq 0}\tilde{\beta}_{k} (x_{m_0})\lambda_0^{k+2} ,\\
    \label{gamma m RG zero}
    &\tilde{\gamma}_m(\lambda_0,x_{m_0}) \equiv \mu\frac{\partial \text{ln}\;m_0^2}{\partial \mu}\Big|_{x_{m},\lambda=\text{fixed}} = \sum_{k\geq 0}\tilde{\gamma}_{m,k} (x_{m_0})\lambda_0^{k+1},\\
    &\tilde{\gamma}_n(\lambda_0,x_{m_0}) \equiv \mu \frac{\partial  \text{ln}\tilde{Z}_n}{\partial \mu}\Big|_{x_{m},\lambda=\text{fixed}} = \sum_{k\geq 0}\tilde{\gamma}_{n,k} (x_{m_0})\lambda_0^{k+1},
\end{align}
\end{subequations}
with the renormalization scheme-independent coefficients $\tilde{\beta}_{k} (x_{m_0})$, $\tilde{\gamma}_{m,k} (x_{m_0})$, and $\tilde{\gamma}_{n,k} (x_{m_0})$. Moreover, comparing \eqref{eq:RG betas} and \eqref{eq:RG betas zero}, one has that $\beta_0 = \tilde{\beta}_0$, $\beta_1 = \tilde{\beta}_1$, $\gamma_{n,0}=\tilde{\gamma}_{n,0}$, and $\gamma_{m,0}=\tilde{\gamma}_{m,0}$.

Let us also discuss, that one can follow the logic of \cite{Kataev:2013eta} and choose $Z_{\lambda}=1$, $Z_m=1$, so that the scheme-independent coefficients from \eqref{eq:RG betas zero} become equal to the coefficients of expansion \eqref{eq:RG betas} with fixed bare parameters. The latter effective renormalization prescription forms the infinite parametrized Lie subgroup of the renormalization group, which was considered in Refs.~\cite{Goriachuk:2020wyn,Kataev:2024xbl}. The supersymmetric QED considerations of Ref.~\cite{Kataev:2019olb} give us a hint that the OS scheme enters this mentioned subgroup.

Finally, we note that in \cite{Callan:1975} all considerations are based on the use of \textit{skeleton expansion} of correlation functions, i.e. $\Gamma_0^{(n)}$ are just graphs for which no \textit{subgraphs} with positive degree of divergence are subtracted. This immediately means that the connection between renormalized correlation function and bare one is given by \eqref{R-op-new}, where we have fixed $\mathcal{R}' = 1$, i.e. we do not subtract divergent subgraphs, since they do not arise in skeleton expansion.

\section{The discussion of renormalization conditions} 
Now, we will specify how we define the finite $m(\mu)$ and $\lambda(\mu)$, since these quantities  are not fixed in \eqref{eq:RG betas}, i.e. for now eq.~\eqref{RG} is written in an arbitrary renormalization scheme. To define finite $\lambda$, one may adopt the \textit{momentum-subtraction} scheme at zero external momenta, so that renormalized coupling reads \cite{Collins:1984xc}:
\begin{subequations}
\label{eq:OffShell}
\begin{align}
\label{eq:OffShell coup}
    \Gamma^{(4)} \Big(s+t+u = \frac{4}{3}\mu^2\Big) = i \lambda,
\end{align}
with $s\equiv (p_1+p_2)^2, t\equiv (p_1-p_3)^2,u \equiv (p_1-p_4)^2$ being the Mandelstam variables and they are the sum of incoming and outgoing momenta in three different $s-,t-,u-$ channels, respectively.

The full 2-point correlation function is defined as
\begin{align}
    \Gamma^{(2)}(p^2) = -i\big(p^2 - m^2 + \Sigma(p^2,m^2,\lambda)\big),
\end{align}
where $\Sigma(p^2,m^2,\lambda)$ is a self-energy operator.
The renormalized finite mass may be defined as
\begin{align}
\label{eq:OffShell mass}
    \Gamma^{(2)}(p^2 = \mu^2) = -i(\mu^2 - m^2 ) \quad \text{with}  \quad \Sigma(p^2 = \mu^2,m^2,\lambda) =0,
\end{align}
which is an off‑shell renormalization condition.
Another condition that fixes the field renormalization may be chosen as:
\begin{align}
    \frac{d}{dp^2}\Gamma^{(2)}(p^2) \Big|_{p^2 = \mu^2}  = -i , \quad \text{or} \quad \frac{d}{dp^2}\Sigma(p^2) \Big|_{p^2 = \mu^2}  = 0 .
\end{align}
\end{subequations}

If one chooses $\mu^2 =  m^2$, the mass is defined in the \textit{on-shell} scheme:
\begin{align}
\label{on shell mass}
    \Gamma^{(2)} (p^2 = m^2) = 0.
\end{align}
Any chosen renormalization conditions are correct up to all orders and they define the renormalization factors $Z_m$, $Z_2$, and $Z_4$ from \eqref{eq:all renorm} and \eqref{eq:Z2 Z4}.

Having introduced all the necessary notations and expressions, in the next section we turn to the finite CS method of renormalization. Corresponding system of differential equations will be obtained in a fully renormalized language. In this form we will compare CS equations with the RG equation \eqref{RG}. This will allow us to understand which renormalization scheme the CS method corresponds to.

To make the text clearer for the reader, we would like to emphasize once again that: 
\begin{itemize}
    \item The first notation or object which is under the study in this paper is ``the system of differential equations reminiscent of the Callan–Symanzik equations'' \cite{Callan:1975}. We note, that we can also call this object as ``the system of differential CS equations'', ``the system of differential equations used in finite formulation of QFT'', ``CS equations'', or sometimes we can also refer to this as ``CS approach'' or ``CS method''. This system is given by \eqref{system} with \eqref{eq:renorm betas}.
    \item The second notation is ``renormalization group equation''. We can also call it as ``the standard renormalization‑group equation written in the Callan--Symanzik--Ovsyannikov form'' or just ``RG equation'' and related ``RG approach''. This is given by \eqref{RG} with \eqref{eq:RG betas}. 
\end{itemize}

\section{The relation between the system of differential CS equations and RG equation}
\label{sec:compare}
In this section we rederive the system of differential CS equations from Refs.~\cite{Callan:1975,Mooij:2021lbc} in a fully renormalized language, so that $\beta$-function and anomalous dimensions explicitly include all renormalization factors (which were introduced in a quite general manner in \eqref{G2 bold renor} and \eqref{eq:all renorm}).   The logic of the derivation of the CS equations closely follows Refs.~\cite{Callan:1975,Mooij:2021lbc}.


Let us briefly recall the main idea of CS method \cite{Blaer:1974foy,Callan:1975,Mooij:2021lbc}. The latter is based on the observation that differentiating a Feynman graph with respect to the mass gives a sum of terms in which each propagator in the graph is doubled. This operation lowers the
degree of UV-divergence of the particular graph. In Refs.~\cite{Blaer:1974foy,Callan:1975,Mooij:2021lbc} this operation is called ``$\theta$-operation'' and defined as 
    \begin{align}
    \label{eq:theta def}
    &\Gamma^{(n)}_{\theta,0} \equiv -i \times \frac{d}{dm_0^2}\Gamma_0^{(n)}.
\end{align} 
The definition \eqref{eq:theta def} includes bare $\Gamma_0^{(n)}$. 
The renormalization of the left‑hand side of \eqref{eq:theta def} with $n=2$ is
\begin{align}
\label{eq: g2 theta general}
    \Gamma^{(2)}_{\theta,0} (\lambda_0,m_0) = \tilde{Z}_2\tilde{Z}_{\theta}  \Gamma^{(2)}_{\theta}(\lambda,m), 
\end{align}
where $\tilde{Z}_{\theta}$ is introduced to renormalize $\Gamma^{(2)}_{\theta,0} (\lambda_0,m_0)$, and where we emphasize once again that constants with ``tilda'' like $\tilde{Z}_{\theta}$ relates to the choice of some specific renormalization scheme, see eq.~\eqref{R-op-new}.
However, it is known that the corresponding tadpole diagram for $\Gamma^{(2)}_0$ diverges quadratically.\footnote{The detailed discussion of higher-loop orders in the framework of the CS method can be found  in Ref.~\cite{Mooij:2021ojy} and in section 3.1.3 wherein.} Therefore, one needs to apply two $\theta$-operation on this diagram in order to make it finite. To this end, in the framework of CS approach another object should be introduced \cite{Blaer:1974foy,Callan:1975,Mooij:2021lbc}
\begin{align}
\label{Gamma tt}
    \Gamma_{\theta\theta,0}^{(2)} \equiv - i\times \frac{d}{dm_0^2}\Gamma_{\theta,0}^{(2)}.
\end{align}
The connection between renormalized $\Gamma_{\theta\theta}$ and bare $\Gamma_{\theta\theta,0}$ is given by
\begin{align}
\label{Gamma tt renorm}
    \Gamma^{(2)}_{\theta\theta,0}(\lambda_0,m_0) = \tilde{Z}_2\tilde{Z}^2_{\theta}  \Gamma^{(2)}_{\theta\theta}(\lambda,m). 
\end{align}
By the same logic, one obtains for the four‑point function:
\begin{align}
\label{Gamma t 4 renorm}
    \Gamma^{(4)}_{\theta,0}(\lambda_0,m_0) = \tilde{Z}^2_2\tilde{Z}_{\theta}  \Gamma^{(4)}_{\theta}(\lambda,m). 
\end{align}
Next, we rewrite the derivative appearing in \eqref{eq:theta def} and \eqref{Gamma tt} as follows
\begin{align}
\label{eq:deriv m0}
    \frac{d}{dm_0^2}\Big|_{m,\lambda=\text{fixed}}  = \frac{\partial}{\partial m_0^2}\Big(\frac{m_0^2}{\tilde{Z}_m^2}\Big) \frac{\partial}{\partial m^2} + \frac{\partial}{\partial m_0^2}\Big(\frac{\lambda_0}{\tilde{Z}_{4}/\tilde{Z}_2^2}\Big)\frac{\partial}{\partial \lambda},
\end{align}
with $\tilde{Z} = \tilde{Z}_2,$ and $\tilde{Z}_{\lambda} =\tilde{Z}_4/\tilde{Z}_2^2$.

Substituting \eqref{eq: g2 theta general}, \eqref{Gamma tt renorm}, and \eqref{eq:deriv m0} into the definitions of $\theta$-operation \eqref{Gamma t 4 renorm}, \eqref{Gamma tt}, and \eqref{eq:theta def}, one arrives at
\begin{subequations}
\label{system}
\begin{align}
\label{CS for 4}
&2im^2\tilde{\gamma}_m\Gamma^{(4)}_{\theta} = \Big[\left(2m^2\frac{\partial}{\partial m^2} + \tilde{\beta}\frac{\partial}{\partial \lambda}\right)+4\tilde{\gamma}_4\Big]\Gamma^{(4)},\\
\label{second CS}
    &2im^2\tilde{\gamma}_m\Gamma^{(2)}_{\theta\theta} = \Big[\left(2m^2\frac{\partial}{\partial m^2}  + \tilde{\beta} \frac{\partial}{\partial   \lambda}\right)+2\tilde{\gamma}_2 +\tilde{\gamma}_{\theta}\Big]\Gamma^{(2)}_{\theta}, \\
\label{new renorm CS}
&2im^2\tilde{\gamma}_m\Gamma^{(2)}_{\theta} = \Big[\left(2m^2\frac{\partial}{\partial m^2} + \tilde{\beta}\frac{\partial}{\partial \lambda}\right)+2\tilde{\gamma}_2\Big]\Gamma^{(2)},
\end{align}
\end{subequations}
respectively, and where we have introduced the finite $\beta$-function and anomalous dimensions with the renormalized 
$m,\lambda$ held fixed:
\begin{subequations}
\label{eq:renorm betas}
\begin{align}
    &\tilde{\gamma}_m \equiv \Big[\frac{\partial}{\partial m_0^2}\Big(\frac{m_0^2}{\tilde{Z}_m^2}\Big)\Big]^{-1} \tilde{Z}_{\theta}\Big|_{m,\lambda=\text{fixed}},\label{eq:mass gamma}\\
    &\tilde{\beta} \equiv  2m^2\Big[\frac{\partial}{\partial m_0^2}\Big(\frac{m_0^2}{\tilde{Z}_m^2}\Big)\Big]^{-1}\frac{\partial}{\partial m_0^2}\Big(\frac{\lambda_0}{\tilde{Z}_{4}/\tilde{Z}_2^2}\Big)\Big|_{m,\lambda=\text{fixed}},\\
    &\tilde{\gamma}_2 \equiv m^2\Big[\frac{\partial}{\partial m_0^2}\Big(\frac{m_0^2}{\tilde{Z}_m^2}\Big)\Big]^{-1} \frac{\partial\; \text{ln} \tilde{Z}_2}{\partial m_0^2}\Big|_{m,\lambda=\text{fixed}},\\
     &\tilde{\gamma}_4 \equiv m^2\Big[\frac{\partial}{\partial m_0^2}\Big(\frac{m_0^2}{\tilde{Z}_m^2}\Big)\Big]^{-1} \frac{\partial\; \text{ln} \tilde{Z}_4}{\partial m_0^2}\Big|_{m,\lambda=\text{fixed}},\\
    &\tilde{\gamma}_{\theta} \equiv 2m^2\left[\frac{\partial }{\partial m_0^2} \Big(\frac{m_0^2}{\tilde{Z}_m^2}\Big)\right]^{-1}\frac{\partial \;\text{ln}\;\tilde{Z}_{\theta}}{\partial m_0^2}\Big|_{m,\lambda=\text{fixed}} \label{gammaTeta}.
\end{align}
\end{subequations}

The system of equations \eqref{system} allows one to derive $\Gamma^{(2)}$ in a fully finite way \cite{Blaer:1974foy,Callan:1975,Mooij:2021lbc}: $\Gamma^{(2)}_{\theta\theta}$ and $\Gamma^{(4)}_{\theta}$ are already finite by construction, $\Gamma^{(2)}_{\theta}$ can be found from \eqref{second CS} and then $\Gamma^{(2)}$ -- from \eqref{new renorm CS}, while $\Gamma^{(4)}$ -- from \eqref{CS for 4}. For completeness, the system of CS equations \eqref{system} must be supplemented by some \textit{ad hoc} boundary conditions, which actually correspond to the choice of concrete renormalization scheme. Indeed, solving equations \eqref{system} with \eqref{eq:renorm betas}, we obtain the answer for $\Gamma^{(2)}$ and $\Gamma^{(4)}$ up to the integration constants, which should be fixed by boundary conditions. Thus, the particular form of $\Gamma^{(2)}$ and  $\Gamma^{(4)}$ depend on the choice of renormalization scheme. Nevertheless, this method works for any choice of external momentum scale at which the boundary conditions are imposed, see discussion in \cite{Mooij:2021ojy,Mooij:2021lbc}. Moreover, the authors of \cite{Naud:1998yg} have also shown that the proof of renormalizability through the use of the Callan--Symanzik equation can be done without imposing any normalization conditions.

\section{The choice of renormalization conditions}
\label{sec:on-shell}
The difference between equations \eqref{system} with \eqref{eq:renorm betas} and what was used in \cite{Callan:1975,Mooij:2021lbc} is as follows. Although eqs. \eqref{system} are exactly the same as the CS equations from \cite{Callan:1975,Mooij:2021lbc} (up to notation), now all $\beta$-functions and anomalous dimensions \eqref{eq:renorm betas} explicitly include the renormalization factors $\tilde{Z}_{\theta}$, $\tilde{Z}_m$, $\tilde{Z}_2$, and $\tilde{Z}_4$, which we have introduced in a general manner. To find these constants and then $\beta$-function and anomalous dimensions in each order of perturbation theory one should stick to some renormalization scheme or conditions to define the renormalized mass and coupling constant, as well as the scale at which they are defined. So, in order to understand which renormalization scheme the equations \eqref{system} with \eqref{eq:renorm betas} correspond to, let us now consider the RG equation \eqref{RG zero}.

As mentioned in Section~\ref{sec:general}, the RG equation \eqref{RG zero} is written in arbitrary
renormalization scheme until the renormalized mass $m(\mu)$ and coupling $\lambda(\mu)$ are defined. Let us define mass $m(\mu)$ and coupling $\lambda(\mu)$ through the conditions \eqref{eq:OffShell}. Moreover, we choose the renormalization scale such that $\mu^2 = m^2$ in \eqref{eq:OffShell mass}, so that we arrive to \eqref{on shell mass}. \textit{This corresponds to the momentum‑subtraction scheme at zero external momenta for the coupling constant and the on‑shell scheme for the mass}.

We now arrive at the main point: under this choice and with $\mu^2 = m^2$, eq.~\eqref{RG zero} with $n=2,4$ becomes term‑by‑term \textit{equivalent} to the system \eqref{system}.\footnote{Since we find $\Gamma_{\theta}^{(2)}$ from equation \eqref{second CS}, then one should compare only one CS equation \eqref{new renorm CS} with RG equation \eqref{RG zero} with $n=2$ within chosen renormalization scheme. Obviously, in the case of four-point correlation function, one should compare \eqref{RG zero} with $n=4$ and \eqref{CS for 4}.} Indeed, consider, for example, the following term from \eqref{new renorm CS} and proceed the mentioned substitutions:
\begin{align}
\label{one term}
    &im^2\tilde{\gamma}_m\Gamma^{(2)}_{\theta} = im^2\tilde{\gamma}_m\frac{\Gamma^{(2)}_{\theta,0} (\lambda_0,m_0)}{\tilde{Z}_2\tilde{Z}_{\theta}} \nonumber\\
    &= m^2\Big[\frac{\partial}{\partial m_0^2}\Big(\frac{m_0^2}{\tilde{Z}_m^2}\Big)\Big]^{-1} \tilde{Z}_{\theta}\Big|_{m,\lambda=\text{fixed}}\frac{(d\Gamma_0^{(2)}(\lambda_0,m_0)/dm_0^2)}{\tilde{Z}_2\tilde{Z}_{\theta}}\nonumber\\
&=m^2\frac{\partial\Gamma^{(2)}(\lambda,m)}{\partial m^2} \Big|_{m,\lambda=\text{fixed}}.
\end{align}
Here we emphasize that the factor $\tilde{Z}_m$ just cancels from the final expression in \eqref{one term}.
The latter is finally equivalent to the $\tilde{\gamma}_m(\lambda) m^2 \frac{\partial}{\partial m^2}\Gamma^{(2)}$ from \eqref{RG zero} with $\tilde{\gamma}_m(\lambda)$ which is given by \eqref{gamma m RG zero} with $\mu^2 = m^2$ substituted; in this scheme, the anomalous dimension of the mass \eqref{gamma m RG zero} from \eqref{RG zero} is a constant. 

By the same logic, this equivalence arises for the rest of the terms in the CS system of equations and RG equations:
\begin{align}
     \left[\mu^2 \frac{\partial}{\partial \mu^2}\Gamma^{(2)}\right]_{\mu^2 = m^2}  &= m^2\frac{\partial}{\partial m^2}\Gamma^{(2)},\\
     \left[\beta\Big|_{m,\lambda=\text{fixed}} \frac{\partial}{\partial \lambda} \Gamma^{(2)}\right]_{\mu^2 = m^2} &=\tilde{\beta}\Big|_{m,\lambda=\text{fixed}}\frac{\partial}{\partial \lambda}\Gamma^{(2)},\\
     \left[\gamma_2\Big|_{m,\lambda=\text{fixed}} \Gamma^{(2)}\right]_{\mu^2 = m^2} &=\tilde{\gamma}_2\Big|_{m,\lambda=\text{fixed}}\Gamma^{(2)},
\end{align}
as well as
\begin{align}
    \left[  \gamma_4\Big|_{m,\lambda=\text{fixed}} \Gamma^{(4)}\right]_{\mu^2 = m^2} &=\tilde{\gamma}_4\Big|_{m,\lambda=\text{fixed}}\Gamma^{(4)}.
\end{align}
We emphasize that during the explicit substitution of  $\mu^2 = m^2$ into the RG equation \eqref{RG zero} and the subsequent comparison with CS equations \eqref{system}, one should bear in mind, that $\beta$-function and anomalous dimensions are defined differently for RG equation and for the system of CS equations.

\section{Conclusion}
\label{sec:concl}
In this paper we have shown that renormalization-group equation in the Callan--Symanzik--Ovsyannikov form \cite{Callan:1970yg,Symanzik:1970rt,Ovsyannikov:1956fa} is equivalent to the system of differential equations used in the finite‑QFT formulation \cite{Blaer:1974foy,Callan:1975,Mooij:2021lbc}. However, this equivalence becomes explicit only for a specific renormalization scheme. In particular, the on‑shell scheme for the renormalized mass is required. So, we conclude, that within massive $\phi^4$ toy model the on‑shell mass scheme is singled out for the CS method. 

The applications of on-shell schemes for mass are used in many contexts and sometimes these schemes are indeed distinguished for some reason. Such a scheme was used in the case of QED, for example, in \cite{DeRafael:1974iv,Chetyrkin:1980pr}; in supersymmetric QED in \cite{Kataev:2019olb}; and in quantum chromodynamics (QCD), e.g., in \cite{Jegerlehner:1998zg}.  
However, imposing the on‑shell renormalization condition for the mass is subtle in asymptotically free theories such as QCD. In such models the on-shell mass can be defined in a gauge-invariant way by using a specific gauge-invariant procedure analogous to a subtraction scheme at zero momenta (or at a specific subtraction point). However, calculations then become difficult; see \cite{Jegerlehner:1998zg} for details. From this perspective, implementing the finite‑QFT formulation for QCD requires additional care. 

Important remark to be done, is that in this paper we consider the case of massive scalar theory \eqref{L}. For now, the CS method is only applicable to the massive fields. It would be interesting to search for the generalization of this method to a more general class of theories involving the massless fields as well.

Finally, as mentioned above, 
the coefficients in the expansion for beta-function and anomalous dimensions \eqref{eq:RG betas zero} (which appear in CS method as well) do not depend on the renormalization scheme, while for $\beta$-function and anomalous dimensions \eqref{eq:RG betas} (which appear in RG equation) the situation is the opposite and the coefficients depend on the particular scheme \cite{Itzykson:1980rh,Collins:1984xc}. So, it would be interesting to further explore these scheme (in)dependencies. Moreover, we expect, that  within massive $\phi^4$ toy model and in the framework of \textit{OS scheme for the renormalized mass} the coefficients in expansions \eqref{eq:RG betas} and \eqref{eq:RG betas zero} become equal for the beta-function starting from the third term with $k\geq2$ and for the anomalous dimension starting from the second term with $k\geq1$. This would finally prove the equivalence of the two approaches in the OS scheme for the renormalized mass. Within the $\phi^4$ toy model, this requires computing the anomalous dimension at two loops, the $\beta$-function at three loops, and the Green functions up to the scheme depending level; the results of, e.g., Refs.~\cite{Belokurov:1973fi,Batkovich:2016jus,Kompaniets:2017yct} may be useful for further research.

\section*{Acknowledgments}
The authors are grateful to D. I. Kazakov, N. V. Krasnikov, M. V. Kompaniets, and in particular to M. E.  Shaposhnikov for useful comments and fruitful discussions. The work of YA has been supported by Russian Science Foundation Grant No. 24-72-00121, https://rscf.ru/project/24-72-00121/.

\end{document}